\input stromlo

\def\mathmode#1{\ifmmode {#1} 
                  \else {$#1\mkern-5mu$} \fi}

\def\feh{{\rm [Fe/H]}}
 
\def\mlam{\mathmode{m_{\lambda}}}

\def\msun{\mathmode{M_\odot}}
\def\lsun{\mathmode{L_\odot}}
\def\mlam{\mathmode{m_\lambda}}
\def\xvv{\hbox{$15\!-\!V$}}

\def\et{et~al.}
\def\teff{\mathmode{T_{_{\rm eff}}}}
\def\wrange#1#2{\mathmode{ {#1} < \lambda < {#2} \, {\rm \AA}}}
\def\wless#1{\hbox{\mathmode{\lambda < {#1} \, {\rm \AA}}}}

\def\gta{\mathmode{\geqsim}}
\def\lta{\mathmode{\leqsim}}
\def\mgii{\mathmode{\rm Mg_2}}
\def\xvv{\mathmode{(15-V)}}
\def\solar{\mathmode{_\odot}}
\def\etal{{\it et~al.}}
\def\kelvin{\mathmode{\, {\rm K}}}

\title Ultraviolet Light from Old Stellar Populations

\shorttitle UV radiation from Old Stars

\author Ben Dorman^{1,2}

\shortauthor Dorman

\affil @1 Laboratory for Astronomy \& Solar Physics, NASA/GSFC

\affil @2 Department of Astronomy, University of Virginia

\abstract 

We consider the general theoretical problem of ultraviolet light from
old stellar populations ($t \, \gta 2 \, {\rm Gyr}$), and the
interpretation of galaxy spectra at short (\wless{3000}).  The sources
believed to be responsible for the observed `ultraviolet upturn
phenomenon' (UVX) are Post-AGB (P-AGB), extreme horizontal branch (EHB)
and AGB-Manqu\'e (AGBM) stars. All of these have observational
counterparts in old Galactic stellar populations, i.e. in globular
clusters and in the Galactic field. The production of EHB stars depends
on mass loss on the red giant branch, which is poorly understood,
making the far-UV flux problematical as an indicator of the gross
properties, such as the age, of a given galaxy. On the other hand, the
longer, mid-UV wavelengths are radiated by stars near the main sequence
turnoff which has well understood and predictable behavior.
We discuss the current state of the comparison between theory and
observation for the \index{UVX phenomenon} UVX phenomenon, and revisit the interpretation
of the well-known \index{UVX-\mgii\ correlation}
UVX-\mgii\ correlation. In particular, we draw attention
to the fact that the UVX appears not to be correlated with indices
that measure the iron abundance, which has implications for models
that explain the UV-\mgii\ correlation as an abundance-driven effect.
Finally, we note  the potential utility of ultraviolet observations in
distinguishing age and metallicity from galaxy spectral energy
distributions.

\section Introduction

The ultraviolet upturn (often referred to as the ``UVX phenomenon'')
from E/S0 galaxies and spiral galaxy bulges was
first observed by the OAO-2 satellite in the late
sixties (Code 1969). It consists of a far-UV (\wless{2000})
 component in the spectra of all such galaxies so far
observed. It is found to have similar radial structure to the optical
light profile, and its amplitude is positively correlated with the \mgii\ line
strength and with the velocity dispersion.  This `UVX component' of the
population is now thought to consist of stars similar to the hot
subdwarf sdB/sdO stars found in the Galactic field and some globular
clusters. The characteristics of these stars coincide with those of
theoretical models of extreme horizontal branch (EHB) and ``AGB-manqu\'e''
stars.  The former  are core--helium-burning objects with very small
hydrogen rich envelopes. The latter are their  hot post-HB
helium-burning progeny that have insufficiently massive envelopes to
reach the asymptotic giant branch after the core helium supply is
exhausted.

It is convenient to define `far-ultraviolet' to refer to the
spectral range $\wrange{912}{1800},$ and `mid-ultraviolet'
to mean $\wrange{1800}{3000}.$ The observations are determined
by the available detector technology: Caesium Iodide photocathodes 
used for the far-UV range have a sharp response cutoff at 1800\AA,
while the mid-UV range detectors use Caesium Telluride which has
little response below 3200\AA.

From the theoretical point of view, in old populations with turnoffs of
spectral type F5 or later there is a strong break in the continuum
around 2000 \AA\ in synthetic spectral energy distributions (SEDs) for
metallicities $\feh \gta -0.5$ 
(see  \S 2.2).  Thus, the SEDs  of galaxies are
generally composite in the mid-UV range, with the flux
originating from the main-sequence (MS) and subgiant branch (SGB) and
from hot, highly  evolved stars that may be  present.  
As we shall argue in \S 3, the mid-UV spectral range can in principle be
decomposed into its two major contributors. Since the turnoff light is
very sensitive to age and metallicity, the proper measurement and
interpretation of this wavelength range holds promise as a source of
population indicators for passively evolving elliptical galaxies.

This review is organized as follows. \S 2 describes
the stellar sources of ultraviolet radiation in old stellar populations,
both in the far- and the mid-UV. \S 3 discusses the comparison between
theoretical expectations and some of the observations; further
details of the observational record can be found in Dorman, O'Connell,
\& Rood 1995 (hereafter DOR95) and references therein. \S 4 briefly
summarizes  observations from the Galaxy and globular cluster system
pertinent to understanding the UVX phenomenon, while \S 5 discusses
the UVX phenomenon and the correlation with \mgii. The last section
summarizes the potential for age/metallicity diagnostics in the mid-UV.

\section Theoretical Background

\subsection Horizontal Branch Morphology and
the Production of Far-Ultraviolet Flux

The sources of far-UV radiation in old stellar populations that are
most significant are those which are powered by stable nuclear
reactions, i.e. hydrogen shell, helium core and helium shell nuclear fusion
(for a discussion of other possible sources, see Greggio \& Renzini 1990).

Evolutionary paths for the various types of hot objects are 
illustrated in Fig. 1. Further discussion on the
details of the evolution can be found in
Dorman, Rood, \& O'Connell (1993, hereafter DRO93)
Castellani \& Tornamb\`e (1991), and Brocato \etal\ (1990)
amongst others.

\figureps[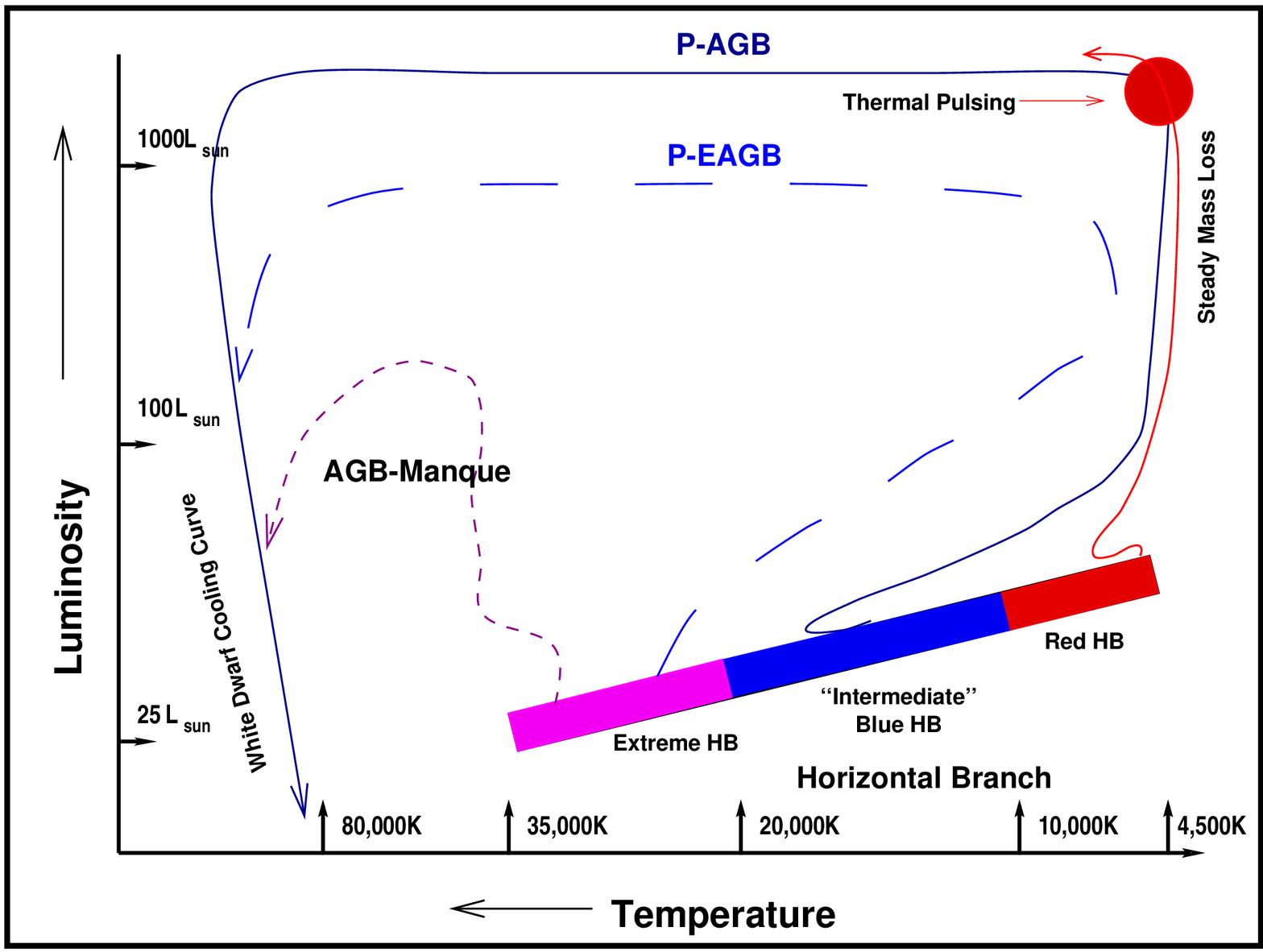,.9\hsize] 1. Schematic of the far-UV emitting
stars in old stellar populations. 

\item{{\bf a)}} {\bf Post-AGB stars:}
This is the `classical' path for AGB stars crossing to the white dwarf
cooling curve, from $\teff \sim 3000$ to $\sim 10^5 \kelvin.$ The lifetime
in this phase ranges from a few times $10^4$ yr to a few thousand years,
decreasing strongly with the luminosity. The spectrum integrated over this
phase has a high characteristic temperature, and its total energy decreases
with increasing P-AGB luminosity.

\item{{\bf b)}} {\bf Hot HB stars:}
Stars that reach the HB with $\teff \, \gta \, 10,000 \kelvin$
can be formed for any metallicity, provided sufficient mass loss
occurs on the red giant branch. Their lifetime
is substantial, $\rm \sim 10^8\, yr,$ with luminosities
$\sim 10-50 \lsun.$ For high metallicity nearly the entire mass
of the envelope must be stripped in order to produce a hot star
(see also D'Cruz \etal\ 1996), while at low $Z$ models with
substantial hydrogen-rich envelopes can be hot enough to radiate
strongly in the far-UV. For sufficiently young systems ($t \sim
5\, {\rm Gyr}$) the amount of mass loss required to produce hot stars
is much greater, so that hot HB stars are expected to be rare or not
present. This qualitative statement unfortunately has no theory of
mass loss available to make it quantitative.

\item{} After the core helium is exhausted, helium shell-burning is initiated.
If the remaining envelope at this stage is sufficiently large ($> 0.05
\msun$) the hot HB stars return to the AGB (i.e., they develop a
convective exterior) as do their more massive red HB counterparts.
They evolve through the thermally pulsing stage, finally becoming P-AGB
stars and white dwarfs.
Otherwise, they become Post-Early AGB stars or
AGB-Manqu\'e stars and are termed EHB stars. The EHB stars have
envelope masses $\lta 0.05 \msun,$ generally have insignificant
hydrogen burning energy, and their effective temperatures range from
about 20,000 to 37,000~K.

\vfil\eject

\item{{\bf c)}} {\bf AGB-Manqu\'e and P-EAGB Stars:}
The core He-burning progeny of stars which have undergone extreme
mass loss, these stars contribute more UV flux in the post-HB
phase than do the classical P-AGB stars. The P-EAGB  (Post-Early
AGB) stars reach the ``early'' ascending AGB stage but not the
thermally pulsing regime at $\log L \, \gta \, 3.2.$
They cross the HR diagram on tracks parallel to the P-AGB models,
but at lower luminosity and with a characteristic timescale of
1 Myr. The AGB-Manqu\'e stars are hot throughout the He-shell burning
phase, with lifetimes typically $\rm 2-4 \times 10^7 \, yr$
(similar to the AGB itself), with $L \sim 100-500 \lsun.$

We stress that for HB stars  {\it the total lifetime energy output
varies by less than 50\% for any abundance.} This is because the mass
of the helium core---controlled by the He flash at the RGB tip---varies
relatively little with abundance.  Thus the helium core luminosity
during the HB phase also varies little. The largest changes with
composition occur with high helium enhancement ($Y \,\gta\, 0.40$), in
which the hydrogen burning shell is hotter for the same envelope mass
(Dorman 1992). The hydrogen burning luminosity is significant for
a larger range in masses and the hydrogen energy production is
greater.  In contrast, the effect of including high $Z$ values in the
models {\it with no concomitant increase in $Y$}  works to decrease
slightly both the mass range and total flux radiated during the EHB lifetime.

Because of this relative invariance in total UV output, it is possible
to constrain the UV-bright population rather tightly. We can regard
UV/Optical colours such as $\mlam(1500) - V$ [hereafter \xvv] as
indicators of the specific frequency of UV sources  (the number present
per unit luminosity). 
It also allows us to compare the UV properties of
populations whose metallicities are vastly different but whose
content may well be similar in age.
The possibility of high $Y$ abundance, which we
have discussed elsewhere (DOR95),  only adds at most a factor of two
uncertainty in the number fractions of EHB stars deduced to be present
(Dorman 1997a).

\subsection The Composite Nature of the mid-UV Radiation

The major contribution to the mid-UV range apart from the hot stellar
component is from the stars close to the main-sequence turnoff, which
is the next hottest population in a quiescent system. The importance of
this spectral range and its potential appears to have first been
stressed by Burstein \et\ (1988), who noted that, if the UVX could be
understood {\it ``... the 2600--3000 \AA\ spectrum may ultimately
provide better constraints on age and composition than are available
now from ground based data.''}

Figure 2 taken from Dorman \& O'Connell (1996) shows the contributions
from various parts of an old stellar population at wavelengths
\hbox{\wless{4000},} normalized to the $V$-band.  In this diagram, the
pre-He-flash stages of evolution (the ``isochrone'' contribution) is
divided into three parts: the lower main sequence, with $\teff < 5000
\kelvin$ and $L < L_{\rm TO};$ the turnoff region,  with $\teff > 5000
\kelvin,$ which includes a portion of the main sequence and of the
subgiant branch; and lastly the RGB itself.  The synthetic spectra in
this figure assume a 10 Gyr population, solar metallicity, and a
Salpeter IMF.  The contribution from lower main sequence
stars (not shown) is similar to that of the red HB stars.  The
possible hot components are represented by the integrated spectrum of
an EHB star that evolves to an AGB-Manqu\'e sequence (upper curve) and
of a P-AGB star. These spectra are computed by assuming the entire
post-RGB population  passes through this evolutionary phase, and are
weighted by the lifetime and the `specific evolutionary flux,' defined 
as the
number of stars entering the UV bright phase per unit time per unit
magnitude of the parent population. The population synthesis scheme
used here is described in DOR95.

\figureps[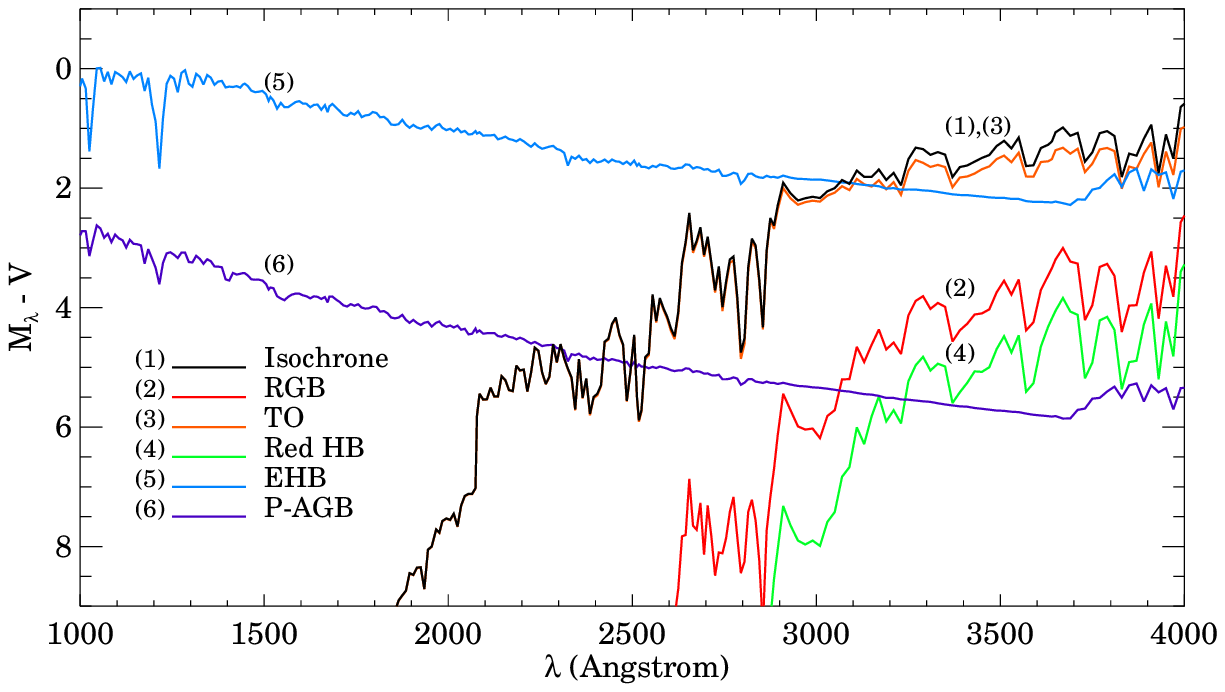,.9\hsize] 2. The components of an old stellar
population for \wless{4000}. The ``isochrone'' component (1) is split
into 3 parts: the red giant branch (2; $\teff < 5000 \kelvin,$ $L >
L_{\rm TO}$), the turnoff region, (3; $\teff > 5000 \kelvin,$) and the
lower main sequence, which is similar in flux to the red HB (4). The
maximum possible contribution from EHB stars (including their
AGB-Manqu\'e progeny) (5)  and P-AGB stars (6) are also shown. The
models of late evolutionary stages are from DOR93, and the earlier
stages were computed for the DOR95 study, together with synthetic
fluxes from Kurucz 1992. Integrated fluxes are computed as described
in  DOR95.

Figure 3 shows the variation of mid-UV flux with metallicity. Plotted are 
spectra from 10 Gyr isochrones for $-2.26 < {\rm [Fe/H]} < 0.58$ with no
synthetic hot component added. For comparison we show the observed spectra
of the M31 bulge from IUE to which has been added a ground-based observation
through a matched aperture (D. Calzetti, private communication; see also
Worthey, Dorman \& Jones 1996), and the Astronomical Netherlands Satellite
(ANS) flux for the metal-poor globular cluster M13 (van~Albada, de~Boer,
\& Dickens 1981). The observations show that the difference in metallicity
produces a large change in the flux level.

\figureps[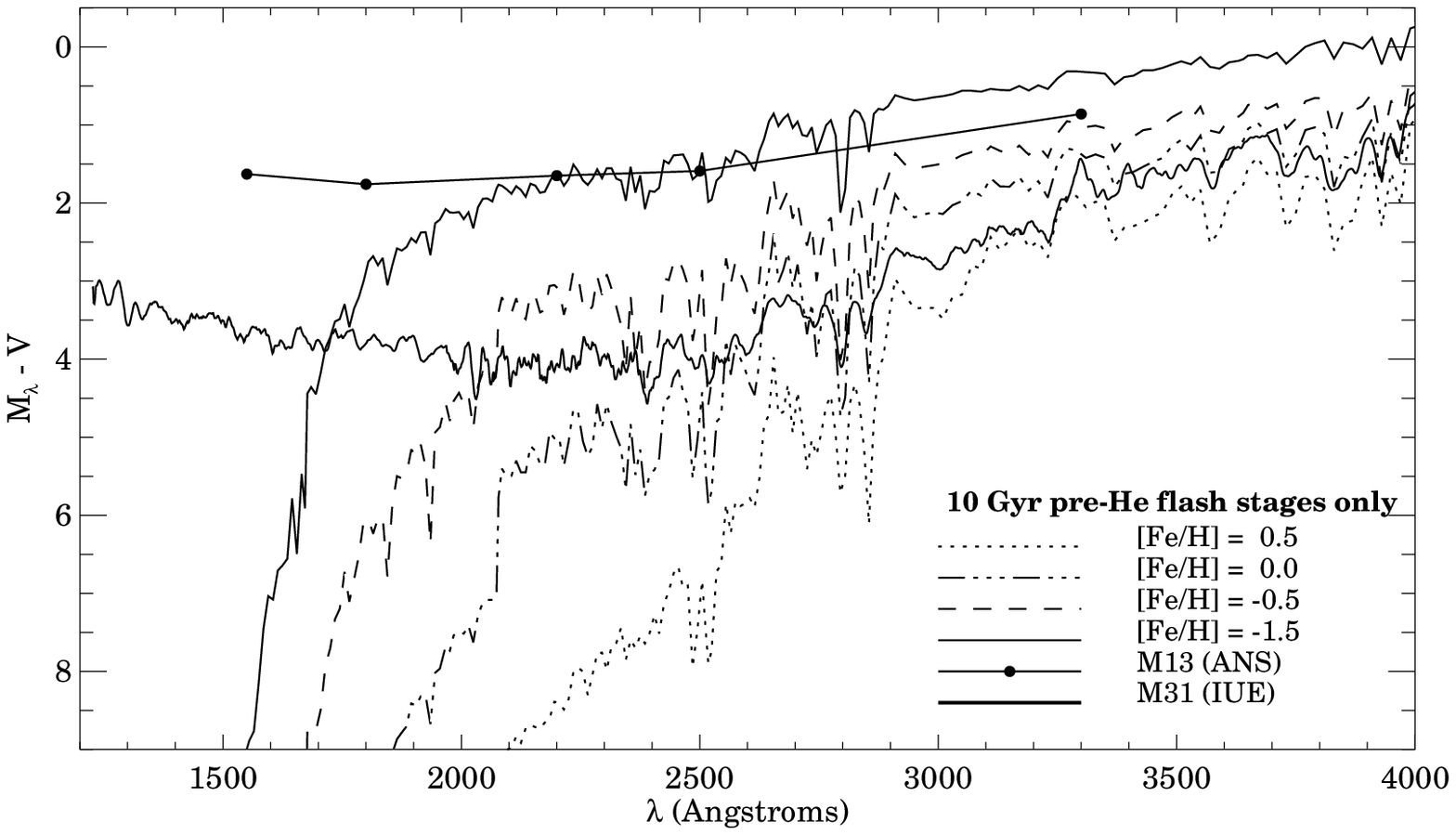,.9\hsize] 3. The variation with metallicity of the
mid-UV spectral range. The models are computed as in figure 2, for 
$-2.26 < {\rm [Fe/H]} < 0.58$ at 10 Gyr. Also plotted are the
IUE + optical spectrum of M31 (see Worthey, Dorman, \& Jones 1996) and the
ANS spectrum of M13.

\section Summary of Theory and Observation

Synthesis models constructed with isochrones plus astrophysically consistent
horizontal branch populations of different morphologies (DOR95, Dorman \&
O'Connell 1997) predict various features of the UV radiation (\wless{3200}).
First, the strength of the far-UV component
is proportional to the number of hot stars that are produced.
 In Fig.~2, the SED generated by the EHB component
produces a  bluest possible 
\xvv\ colour close to 0. Since what we observe is 
$\xvv\, \gta\,  2,$ it follows that the fraction of
EHB stars compared to the total HB population is $\lta\, 15\%.$ 
For solar or greater metallicity the rest of the HB stars will 
lie at the red end  of the HB and eventually become P-AGB stars.
Second, the EHB assumption implies a characteristic temperature for
the UVX in the range  $\sim 20,000 < \teff < 37000 \kelvin.$ Brown,
Ferguson \& Davidsen (1995 and these proceedings) deduced characteristic temperatures
from Hopkins Ultraviolet Telescope data  between 20000 and 23000 K,
and obtained the best model fits to the far-UV spectrum with a small
range in EHB masses. The UV radiation in the galaxies is also observed
to be radially concentrated, like the background population (O'Connell \etal\
1992; Ohl \& O'Connell 1997). Both of these imply an origin for the UV
upturn in the background stellar population rather than due to young
stars in discrete star forming regions.

The maximum possible theoretical contribution from P-AGB stars
or P-EAGB stars gives insufficient UV flux [$\xvv \sim 3.6$] to be
responsible for the strongest UV upturns observed.  
In addition, since
the P-AGB flux depends strongly on the mean luminosity of the AGB tip
only, it is decoupled from the EHB output. A possible indicator of the
relative importance of each may in principle be derived from the
strength of the Lyman absorption lines which are much stronger in the
cooler EHB sequences.

In the mid-UV, the contribution from the red giants is insignificant
shortward of $\sim 3500 {\rm\, \AA},$ and for \wless{2000} the hot
stars are the only contributors.  Hot stellar spectra are, however,
relatively featureless in the mid-UV range. Thus the spectrum of a
galaxy with a UV rising branch will have the `signature' of the
background turnoff population superposed, including the spectral
features appropriate to a late F/early G type population.
The effect of the hot component on the mid-UV may therefore in
principle be quantified and subtracted to yield population indicators
(Burstein \et\ 1988; Dorman \& O'Connell 1996).

Because the mid-UV radiation from metal-poor populations is so much
greater than for the metal-rich, they will have shallower ultraviolet
slopes, as  observed in the flatter UV spectra of globular
clusters and in their measured $(15-25)$  colours (DOR95).  The ratio
of mid-UV to optical flux can therefore be used to constrain the
metallicity spread in the population. 
The mid-UV colours of galaxies [$(25-V) \, \gta\, 3.5$] appear
to rule out fractions of metal-poor stars (i.e. [Fe/H] $\le -1$)
greater than about 10\% since the UV continuum of the isochrone 
component is at $(25-V) \sim 2,$ while at least 50\% of the 
galaxy mid-UV continuum will be contributed  by the hot component
in a strong UVX galaxy.
Further, for sufficiently old populations
the short wavelength cutoff is a function of
metallicity rather than age. Hence presence of significant flux
shortward of 2000 \AA\ in excess of what is implied by a hot component
would suggest the presence of metal-poor stars.

Note that a relatively recent starburst, a few times $10^8$
years old, could seriously affect the interpretation of the mid-UV flux.
This possibility must also therefore be investigated: Worthey, Dorman
\& Jones (1996) discuss a potential indicator for any A star population.
 However,
the mid-UV continuum level must give a lower bound to the age,
since the presence of hot contaminants
will make the spectral energy distribution appear younger
than it is. This has immediate application in interpreting observations of 
rest-frame mid-UV continua of galaxies at significant redshift (Dunlop \etal\
1996). 

\section  Old Galactic Stellar Populations and the Production of Hot Stars

Perhaps the strongest motivation for thinking that galaxies contain
populations of EHB stars is that they appear to be a regular
constituent of old stellar populations in our galaxy, in both the
globular clusters and in the Galactic old disk population. This
evidence is reviewed in Dorman (1997b). However, in many
cases---particularly in the Galactic field, but also in some globular
clusters---the HB population is not unimodal, and the most extreme blue
stars in particular are separated from their cooler counterparts by a
gap in temperature. Since the mass difference between cool ($\teff \sim
5000 \kelvin$)  and hot stars in metal rich populations is
theoretically only a  few times  $0.01\, \msun,$ 
the UV flux can vary enormously as a result of small changes in the
total mass loss. The presence of gaps in the HB  sequence, where they
are statistically significant, either implies that more than one mass
loss process is effective, or that a single process may lead to a
bifurcation in HB properties.  Thus the
\xvv\ colour of  old stellar populations is {\it not} a simple function
of age and metallicity.  Models that attempt to derive ages from \xvv\
colours assume simple analytical formul\ae\ for the mass loss such as
the Reimers (1975) ``law,'' and that HB morphology becomes
unambiguously bluer mainly as a result of age. Such models have been
computed by Bressan \etal\ (1994); Park \& Lee (1997) and Yi (1996).
However, these models cannot provide {\it necessary} constraints on the ages
of galaxies without a better understanding of the idiosyncrasies of
stellar mass loss and the apparent variations between different stellar
systems.

\section The UVX--Mg$_{\bf 2}$ Correlation Revisited

Since its discovery, the UVX phenomenon has continued to be a mystery
with relatively few explanatory clues. While we have a viable
hypothesis for the stars responsible for the phenomenon, we do not yet
understand why the UVX phenomenon has the properties that it does. The
most significant insight that has been found into the causes of the UVX
was the discovery by Faber (1983) of its positive correlation with
metallicity-sensitive indices.  Burstein \etal\ (1988) then reported
that this correlation was indeed present on the largest  (and still
today the most coherent and comprehensive) dataset on UV spectra of
early-type galaxies. Burstein \etal\ noted the correlation between the
UVX and the Lick optical absorption line index \mgii.  They
also noted a somewhat looser correlation with the galaxies'
central velocity dispersions $\sigma.$ However, the tightest
correlation was with the \mgii\ index, which was thus held to be
causal. Much theoretical work (Greggio \& Renzini 1990; Horch, Demarque
\& Pinsonneault 1992; Bressan, Chiosi, \& Fagotto 1994; Tantalo
\etal\ 1996, Yi 1996) has attempted to explain the apparent correlation 
with $Z.$

The \xvv\ vs \mgii\ diagram is reproduced in Fig.~4, which is adapted from
DOR95 and includes for
comparison the data from metal-poor Galactic globular clusters on the same axes.
This diagram compares the 
UV flux from globulars whose UV sources we can resolve, measure, and thus
use to constrain the theory with those of the galaxies. 
To our knowledge, the UV flux in the globulars shown is dominated by hot
HB stars.
The galaxies' specific UV flux, directly measured by the UV/optical colour,
is seen to be of the same order of magnitude---albeit weaker---than
that of the bluest clusters in which all of the HB stars
contribute to the far-UV flux. 

\figureps[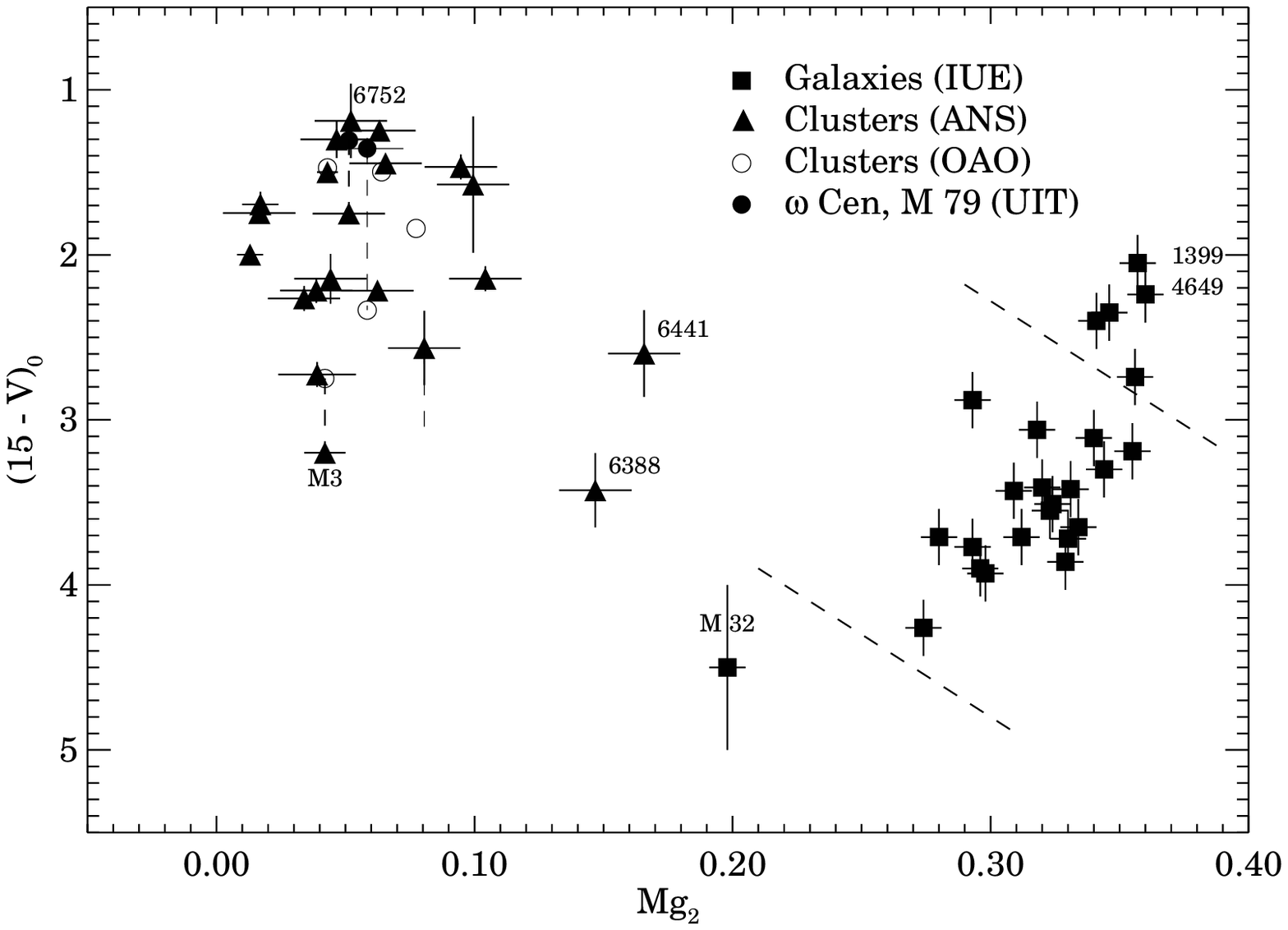,.9\hsize] 4. \xvv\ colours of galaxies and
 Galactic globular clusters. The data are from the compilation in DOR95.
 Dashed lines indicate the division into three regimes referred to in the text.
 NGC numbers for some significant galaxies and clusters are shown.

Explaining the UVX--\mgii\ correlation by invoking some
feature of stellar evolution {\it that is more effective
in producing hot stars at high $Z$} is attractive, 
but requires circumspection.
DOR95 suggested that the UVX--\mgii\ correlation 
actually consists of three different behaviours: (a) ``UVX-weak,'' of
which M32 is so far the lone example. M32 is a Local Group dwarf elliptical
with a mid-UV spectrum that possibly indicates
 a sizeable intermediate age population
 (but see Grillmair \etal\ 1996); (b) the bulk of the ellipticals, SO
types and the spiral bulges that have $\xvv\ = 3-4;$ and (c) the strong
UVX systems, which are all giant ellipticals (we have included the slightly
redder NGC 4889 in Coma as its measured UV/Optical color is probably
underestimated compared to the nearer Virgo and Fornax objects
since more of the galaxy is enclosed in the observing aperture).
The correlation between the UVX and the line index is still present in 
group (b), but the relative scatter is large.

After the landmark paper by Burstein \etal\ appeared it became
well- established that giant elliptical galaxies are overabundant in
$\alpha$-capture elements, including magnesium (Worthey, Faber, \&
Gonzalez 1992 and see the review in Davies 1996).  The \mgii\ index,
strongly correlated with [Fe/H] in the globular clusters (Burstein
\etal\ 1984) therefore does not serve as an indicator of Fe-peak
elements in the super-solar metallicity regime.  Also, the \mgii\ index
is found to be correlated with $\sigma,$ whereas the same is not so for
line indices such as Fe5270, Fe5330 that are designed to measure the
iron abundance (Worthey 1996; Fisher, Franx, \& Illingworth 1995). 

Taken together, the implication is
that there is no strong correlation between iron abundance and the UV
upturn. This may  be a significant factor in the theoretical modelling.
As noted earlier, the potential increase in the ease at which EHB stars
can be produced with increasing metallicity is, in fact, due to using
high values of the helium abundance in the stellar models.  The
assumption of high $Y$  follows from adopting $\Delta Y/\Delta Z\,
\gta\, 2.5$ for $Z > Z\solar.$ If additional data confirms that the UVX
has no strong correlation with Fe indices these models
can only be maintained if $Y$ increases instead with the $\alpha$ 
elements\note{Note also that $\Delta Y/\Delta Z$ is determined
largely from oxygen rather than  iron abundances.}.
Finally, it should be recalled (DOR95) that there is little evidence
for the behaviour of $Y$ with $Z$ {\it in the metal-rich regime;}
extrapolations from the slopes used in attempting to determine the
primordial $Y$ abundance may introduce gross overestimates of $Y$ in
the models.

These observations may be pointing towards a different understanding
for the origin of the UVX. If we can extrapolate from the Galaxy,
EHB stars are a normal constituent of old stellar populations.
Their incidence is apparently higher in the cores of large ellipticals,
perhaps those with distinct isophotal characteristics (Longo \etal\  1989;
see also Merritt, these proceedings). 

While we are arguing that there
is no direct interpretation of the far-UV component, it should also be
stressed the UVX may be used empirically to help in the understanding of
the integrated light at longer wavelengths. Observations in the 
rest-frame far-UV will constrain the effect of any hot component present
and allow the measurement of the mid-UV continuum level that is 
sensitive to age and metallicity. Of course, these properties are 
difficult to distinguish cleanly. In the final section we briefly
mention the potential that the mid-UV holds for insight into
this well-known problem of separating them.

\section The Age/Metallicity Degeneracy and the Ultraviolet

The \index{age-metallicity degeneracy}
`age-metallicity degeneracy' problem refers to the difficulty of
separating age and metallicity effects between stellar populations.
Both  increasing $Z$ and increasing age decrease the mean temperature of 
the stellar population. Further, RGB stars which contribute at least
half the light in the $V$ band are relatively insensitive to both.
Worthey (1994) has stressed that optical broad band colours 
are particularly affected by this problem, and has isolated absorption
line indices that are significantly more sensitive to either age
or metallicity. However, broadband colours have significant advantages
over line indices. They measure the continuum level of a source
over a wider part of the spectrum, which makes accurate synthetic
modelling much easier, and they can be observed to greater distance
with sufficient S/N  to provide usable diagnostics ({\it cf.}
O'Connell 1996).

\figureps[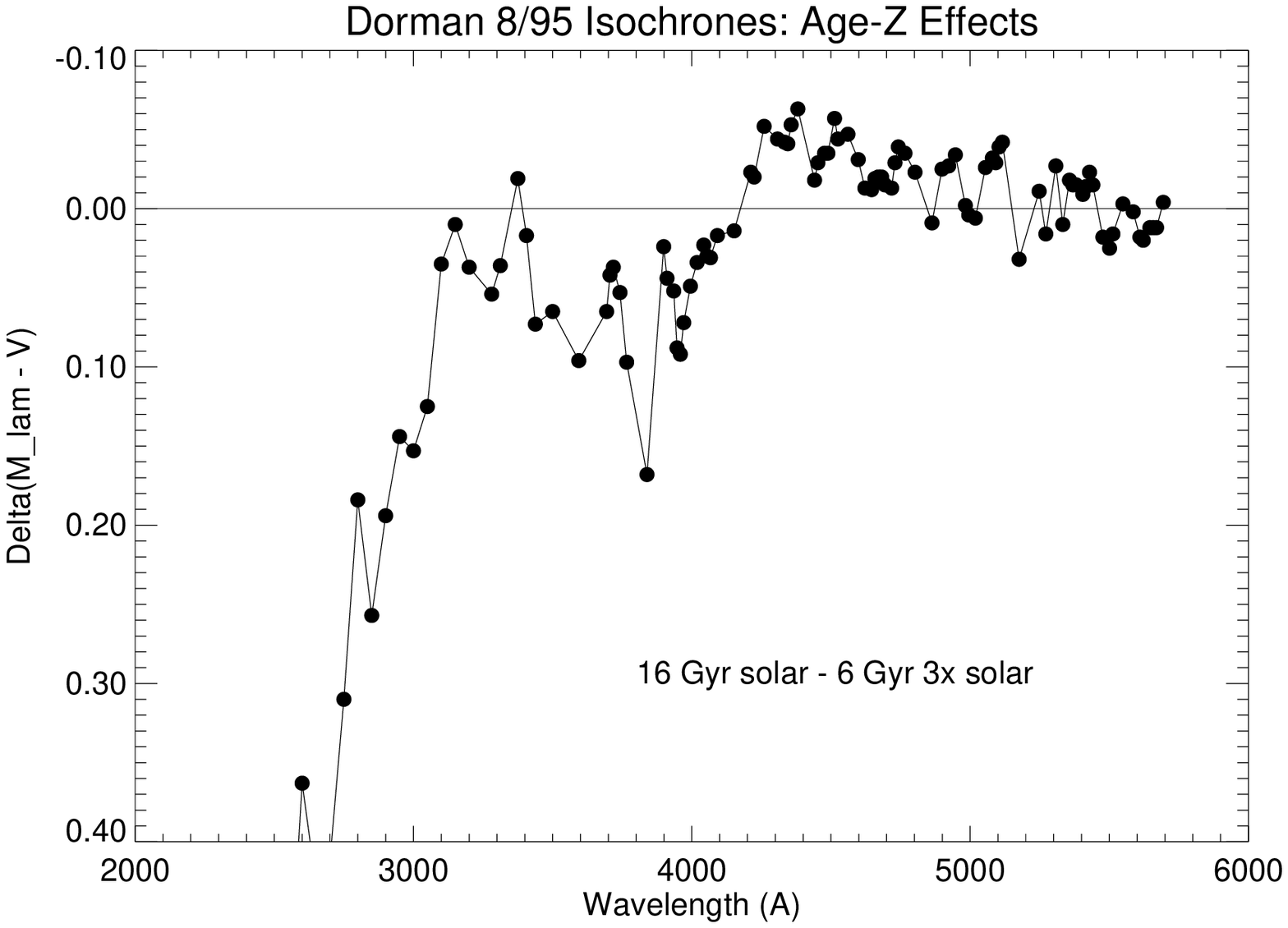,.9\hsize] 5. Comparison between synthetic spectral
energy distributions of `degenerate' models. The difference between
model fluxes computed for $[Fe/H] = 0$ at 16 Gyr has been subtracted from
a 6 Gyr, $[Fe/H] = 0.5,$  and is plotted against wavelength.
The models are from the Crete CD-ROM database
(Leitherer \etal\ 1996) contributed by Dorman \& O'Connell,
described in \S 6.5.

Using UV diagnostics to help resolve this problem has several important
advantages. The mid-UV isolates the turnoff radiation, and thus is
an indicator of the population that varies most strongly with the
gross population properties. A similar goal underlies the use
of H$\beta$ as an age indicator (Gorgas, Efstathiou \& Arag\'on-Salamanca 1990;
Gonzalez 1993).
The spectra of populations that are difficult
to distinguish in the optical, where
the spectra are dominated by red giants, can be very different in the UV. Fig.~5
shows the residuals between two single burst stellar population 
models that are similar at $V$ (they differ by $\lta \, 0.05$ mag)
but are easily distinguished for \wless{4000}.

A number of galaxies have now been studied in the rest-frame mid-UV
from large ground based instruments. The determination of ages
of stellar populations at $z > 1$ is an attractive cosmological tool
because allows the most direct application of the well-established
stellar evolution theory to cosmology. The best known example is
the recent work of Dunlop \etal\ (1996) who obtained a spectrum of the
radio galaxy \index{LBDS 53W091}
LBDS 53W091. Comparison with synthetic spectra imply
a lower bound to the age of 3.5 Gyr. The likelihood that much more
similar data will become available in the future makes the proper
understanding of the mid-UV continuum and its spectral features
an important research goal.

\acknowl I would like to acknowledge the support of the National Research
Council through the Research Associateship programme, 
and partial support from NASA grants NAG5-700 and NAGW-4106.

\references

Bressan, A., Chiosi, C., \& Fagotto, F. 1994, ApJS, 

Brocato, E., Matteucci, F., Mazzitelli, I., \& Tornamb\`e, A., 1990, 
ApJ, 349, 458

Brown, T. M., Ferguson, H. C., Davidsen, A. F., 1995, ApJ, 454, L15

Burstein, D., Bertola, F., Buson, L., Faber, S. M., \& Lauer, T. R., 1988,
ApJ, 328, 440

Burstein, D., Faber, S. M., Gaskell, F., \& Krumm, N. 1984,
ApJ, 287, 586

Castellani, M. \& Tornamb\`e, A. 1991, ApJ, 381, 393 

Code, A. D. 1969, PASP, 81, 475

D'Cruz, N. L., Dorman, B., Rood, R. T., \& O'Connell, R. W. 1996,
ApJ, 466, 359

Davies, R. L. 1996 in {\it New Light on Galaxy Evolution} eds. R. Bender
\& R. L. Davies, (Dordrecht:Kluwer), 37

Dorman, B. 1992, ApJS, 80, 701

Dorman, B. 1997a in {\it Advances in Stellar Evolution} eds. R. T.
Rood \& A. Renzini, (Cambridge:CUP), in press

Dorman, B. 1997b in {\it The Third Conference on Faint Blue Stars}
ed A. G. D. Philip (Schenectady: L. Davis), in press

Dorman, B. \& O'Connell, R. W.  1996 in {\it From Stars to Galaxies:
The Impact of Stellar Physics on Galaxy Evolution}, eds. C. Leitherer,
U. Fritze-von Alvensleben, \& J. P. Huchra (San Francisco:ASP) p. 105

Dorman, B. \& O'Connell, R. W. 1997, in preparation

Dorman, B., O'Connell, R. W., \& Rood, R. T. 1995, ApJ, 442, 105 (DOR95)

Dorman, B., Rood, R. T., \& O'Connell, R. W. 1993, ApJ, 419, 596 (DRO93)

Dunlop, J., Peacock, J., Spinrad, H.,
Dey, A., Jimenez, R., Stern, D., \& Windhorst, R. 1996 Nature, 381, 581

Faber, S. M. 1983, Highlights Astron., 6, 165 

Fisher, D., Franx, M., \& Illingworth, G. D. 1995, 448, 119

Gonzalez, J. J., 1993, Ph.D. thesis, Univ. California -- Santa Cruz

Gorgas, J., Efstathiou, G., \& Arag\'on-Salamanca, A. 1990, MNRAS,
245, 217

Greggio, L. \& Renzini, A. 1990, ApJ, 364, 35

Grillmair, C. J., \etal\ 1996, AJ, 112, 1975

Horch, E., Demarque, P., \& Pinsonneault, M. H., 1992, ApJ, 388, L53

Leitherer, C., \etal\ 1996, PASP, 108, 994

Longo, G. Capacciolli, M., Bender, R., \& Busarello, G., 1989, A\&A, 225, L17

O'Connell, R. W., 1996 in {\it From Stars to Galaxies: The Impact of
Stellar Physics on Galaxy Evolution} eds. C. Leitherer,
U. Fritze von Alvensleben, \& J. P. Huchra (San Francisco: ASP), 3

O'Connell, R. W., \etal\ 1992, ApJ, 395, L45

O'Connell, R. W., \etal\ 1997, ApJ Letters, submitted

Ohl, R. \& O'Connell 1997, in preparation

Park, J. H. \& Lee, Y. W. 1997, ApJ, in press

Reimers, D., 1975, M\'em. Soc. Roy. Astr. Li\`ege, 6$e$ ser. 8, 369

Tantalo, R., Chiosi, C., Bressan, A., \& Fagotto, F. 1996, A\&A, 311, 361

van~Albada, T. S., de~Boer, K. S., \& Dickens, R. J., 1981,
MNRAS, 195, 591

Vassiliadis, E. \& Wood, P. R. 1994, ApJS, 92, 125

Worthey, G. S.,  1994, ApJS, 95, 107

Worthey, G. S. 1996 in {\it From Stars to Galaxies: The Impact of
Stellar Physics on Galaxy Evolution} eds. C. Leitherer,
U. Fritze von Alvensleben, \& J. P. Huchra (San Francisco: ASP), 467

Worthey, G. S., Dorman, B., \& Jones, L. A., 1996, AJ, 112, 948

Worthey, G. S., Faber, S. M., \& Gonzalez, J. J. 1992, ApJ, 398, 69

Yi, S. Y. 1996, Ph.D. thesis, Yale Univ.

\bye